\documentclass{article}
\usepackage{graphics}
\newtheorem{postulate}{Postulate}
\title{Machian General Relativity: a possible solution to the Dark Energy problem and an alternative to Big Bang cosmology }
\author{Robin Booth\\ \small{Theoretical Physics, The Blackett Laboratory} \\ \small{Imperial College, Prince Consort Road, London SW7 2BZ, UK}}

\begin{document}
\maketitle
\begin{abstract}
    Observations of an apparent acceleration in the expansion rate
    of the Universe, derived from measurements of high-redshift
    supernovae, have been used to support the hypothesis that the
    Universe is permeated by some form of dark energy.  We show
    that an alternative cosmological model, based on a linearly
    expanding Universe with Omega=1, can fully account for these
    observations. This model is also able to resolve the other
    problems associated with the standard Big Bang model.  A
    scale-invariant form of the field equations of General
    Relativity is postulated, based on the replacement of the
    Newtonian gravitational constant by an formulation based
    explicitly on Mach's principle.  We derive the resulting
    dynamical equations, and show that their solutions describe a
    linearly expanding Universe. Potential problems with non-zero
    divergencies in the modified field equations are shown to be
    resolved by adopting a radically different definition of time,
    in which the unit of time is a function of the scale-factor.
    We observe that the effects of the modified field equations
    are broadly equivalent to a Newtonian gravitational constant
    that varies both in time and in space, and show that this is
    also equivalent to Varying Speed of Light (VSL) theories using
    a standard definition of time. Some of the implications of
    this observation are discussed in relation to Black Holes and
    Planck scale phenomena. We examine the consequences of the
    definition of time inherent in this model, and find that it
    requires a reappraisal of the standard formula for the energy
    of a photon.  An experiment is proposed that could be used to
    verify the validity of this cosmological model, by measuring
    the noise power spectrum of the Cosmic Microwave Background
    (CMB). An alternative scenario for the early stages of the
    evolution of the Universe is suggested, based on an initially
    cold state.  It is shown that this can in principle lead to
    primordial nucleosynthesis of the light elements.  Finally, we
    examine some of the other potential consequences that would
    follow from the model, including the possibility of
    reconciling the paradox of non-locality inherent in Quantum
    Mechanics, the avoidance of an initial singularity, and the
    ultimate fate of the Universe.

\end{abstract}

\section{Introduction}

\subsection{Big Bang problems}\label{sec:bigbang}
Our ideas about the origin and evolution of the Universe have
developed markedly since the discovery of the Red Shift by Hubble
in 1935.  This phenomenon has lead to the conclusion that the
galaxies must be receding from each other.  This leads to two
possibilities; either the galaxies were once close together and
`exploded' apart in a `Big Bang', or the Universe is in a steady
state, with matter being continually created so as to maintain the
matter density of the Universe at a constant value as it expands.
The subsequent discovery of the microwave background radiation in
1965 suggests strongly that the Big Bang theory is the correct
one. However, the Big Bang theory, whilst very successful in
explaining many features of the Universe including the origins and
relative abundances of the elements, does not fully account for a
number of observations about the large scale structures in the
Universe.  It also leads to disturbing conclusions about the state
of the Universe at times close to the origin of the Big Bang and
at times near the end of the life of the Universe.

The standard cosmological problems are well documented in a number
of sources, including \cite{jm:1}.  A brief summary of the
principal issues is given here.

\subsubsection{The horizon problem}
The standard theory is unable to explain how regions of the
Universe that had not been in contact with each other since the
Big Bang are observed to emit cosmic background radiation at
almost precisely the same temperature as each other.  A related
problem is the explanation for galaxy formation.  Recent results
from COBE and other surveys of the cosmic microwave background
suggest that there were small fluctuations in the distribution of
matter within the Universe at the epoch of radiation decoupling
which could account for the range of large scale structures that
are observed in the Universe.  If the Universe was initially
perfectly smooth, by what mechanism did matter start to clump
together to form stars, galaxies, and galaxy clusters ?

\subsubsection{The flatness problem}
The fact that the observed matter density of the Universe is so
close to the critical value necessary for the Universe to be
closed implies that the ratio of these densities, $\Omega$, must
have been very close to one at the time of the Big Bang. This
observation is the one of the main reasons for the widely held
belief that there is probably additional hidden mass in the
Universe such that $\Omega$ is in fact equal to one in the present
epoch.  The standard theory is unable to provide an explanation as
to why the density of the Universe should be so close to the
critical value.

\subsubsection{The lambda problem}
Einstein originally included the cosmological constant $\Lambda$
in his gravitational field equations in order to arrive at a
solution that was consistent with the prevalent concept of a
static Universe. The subsequent discovery that the Universe is in
fact expanding has done nothing to diminish the enthusiasm on the
part of many theorists for retaining this constant, in spite of
Einstein's opinion that its inclusion was his `biggest blunder'.
Recent measurements of the expansion rate of the Universe appear
to suggest that a non-zero $\Lambda$ may be causing the expansion
to accelerate.  However, the expansion of the Universe would have
caused any initial cosmological constant to grow by a factor of
$10^{128}$ since the Plank epoch. For $\Lambda$ to be as small as
it appears today presents yet another fine tuning problem.

\subsection{Observational evidence for an accelerating
Universe}\label{sec:observations} The expansion history of the
Universe can be explored by measuring the relationship between
luminosity distance and redshift for a light source with a known
intrinsic magnitude. Just such an ideal 'standard candle' has been
identified in the form of Type Ia supernovae. Several studies have
now been undertaken by various groups, including the Supernova
Cosmology Project \cite{perlmutter:1, perlmutter:2} and the High-Z
Supernova Search Team \cite{reiss:supernovae}, to measure the
distances of a relatively large sample of supernovae at redshifts
extending up to $z=0.8$.

\subsubsection{Hubble constant} For low redshifts, the slope of the
Hubble diagram can be used to provide an estimate of the
present-day value of the Hubble constant.  Current measurements of
the Hubble constant \cite{hamuy:1996} give a value of $H_0 = 63.1
\pm 4.5 km s^{-1} Mpc^{-1}$.  Based on this same data, the
standard Big Bang theory would (with $\Lambda =0$)give a value for
the age of the Universe of $t_0 \approx 10 \times 10^9$, which is
at odds with astrophysical and geological evidence.

\subsubsection{Supernovae measurements} For higher redshifts, the
Hubble diagram can tell us whether the expansion of the Universe
has undergone any periods of acceleration in its history. These
results are derived using the expression for the expansion rate in
terms of the redshift
\begin{equation}
 H^2  = H_0^2 \left[ {\Omega _M \left( {1 + z} \right)^3  + \Omega _K \left( {1 + z} \right)^2  + \Omega _\Lambda  } \right] \\
\end{equation}
where
\[
\begin{array}{l}
 \Omega _M  \equiv \left( {\frac{{8\pi G}}{{3H_0^2 }}} \right)\rho _0  \\
 \Omega _\Lambda   \equiv \frac{\Lambda }{{3H_0^2 }} \\
 \Omega _K  \equiv \frac{{ - k}}{{a_0^2 H_0^2 }} \\
 \end{array}
\]
and
\[
\Omega _M  + \Omega _\Lambda   + \Omega _K  = 1
\]

This leads to a formula for the lookback time in terms of the
present value of the Hubble parameter, and the red-shift (see
\cite{goobar} for  derivation)
\begin{equation}\label{lookback}
  t_0  - t_1  = H_0^{ - 1} \int_0^{z{}_1} {(1 + z)^{ - 1} \left[ {\left( {1 + z} \right)^2 \left( {1 + \Omega _M z} \right) - z(2 + z)\Omega _\Lambda  } \right]^{ - {\textstyle{1 \over 2}}} dz}  \\
\end{equation}

For the case of a Universe with $\Omega_M=1$ and
$\Omega_\Lambda=0$ this simplifies to

\begin{equation}
  H_0 t_0  = \int_0^\infty  {(1 + z)^{ - {\textstyle{5 \over 2}}} } dz
\end{equation}

giving the present age of the Universe as

\begin{equation}
t_0  = \frac{2}{{3H_0 }}
\end{equation}

\begin{figure}
  \centering
  \scalebox{1}{\includegraphics*{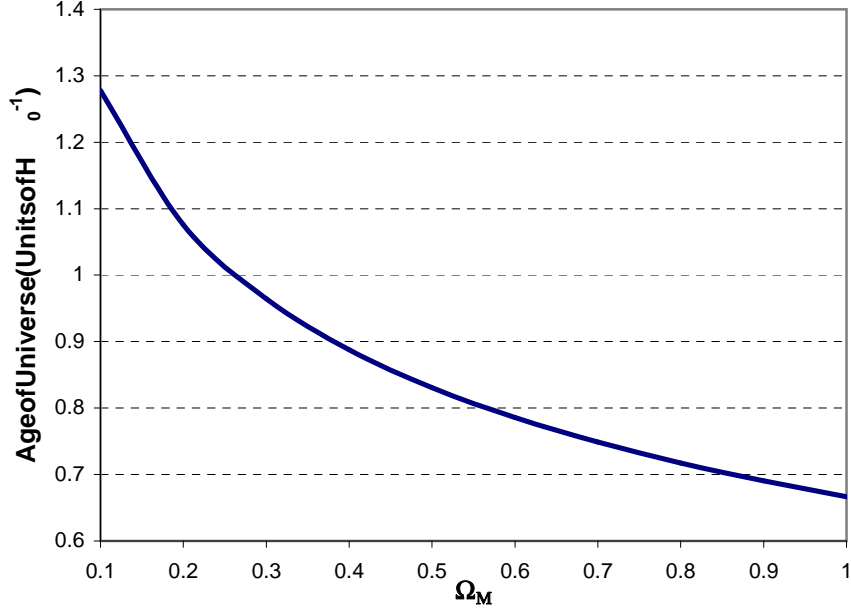}}
  \caption{Age of the Universe as a function of $\Omega_M$, for $\Omega_M + \Omega_\Lambda=1$
  \label{fig:age}}
\end{figure}

Figure \ref{fig:age} illustrates the relationship between the age
of the Universe and $\Omega_M$ for a flat Universe where
$\Omega_{Tot}=1$.

 \par Distance estimates
from SN Ia light curves are derived from the luminosity distance
\begin{equation} d_{L} = \left(\frac{{L}}{4
\pi { F}}\right)^{\frac{1}{2}}
\end{equation}

where ${L}$ and ${ F}$ are the SN's intrinsic luminosity and
observed flux, respectively. In Friedmann-Robertson-Walker
cosmologies, the luminosity distance at a given redshift, $z$, is
a function of the cosmological parameters.  Limiting our
consideration of these parameters to the Hubble constant, $H_0$,
the mass density, $\Omega_M$, and the vacuum energy density (i.e.,
the cosmological constant), the luminosity distance is

\begin{equation}\label{lumdist}
 d_L= c H_0^{-1}(1+z)\left | \Omega_k \right |^{-1/2}{\cal S}\lbrace\left | \Omega_k \right |^{1/2}
\int_0^zF(z'){dz'}\rbrace
\end{equation}

where $\Omega_k=1-\Omega_M-\Omega_\Lambda$, and $\cal S$ is
$\sinh$ for $\Omega_k \geq 0$ and $\sin$ for $\Omega_k \leq 0$,
and

\begin{equation}
 F(z)= [(1+z)^2(1+\Omega_Mz)-z(2+z)\Omega_\Lambda]^{-1/2}
\end{equation}

For $d_L$ in units of Megaparsecs, the predicted distance modulus
is \begin{equation} m-M=5\log d_L +25 \end{equation}

\begin{figure}
  \scalebox{1}{\includegraphics{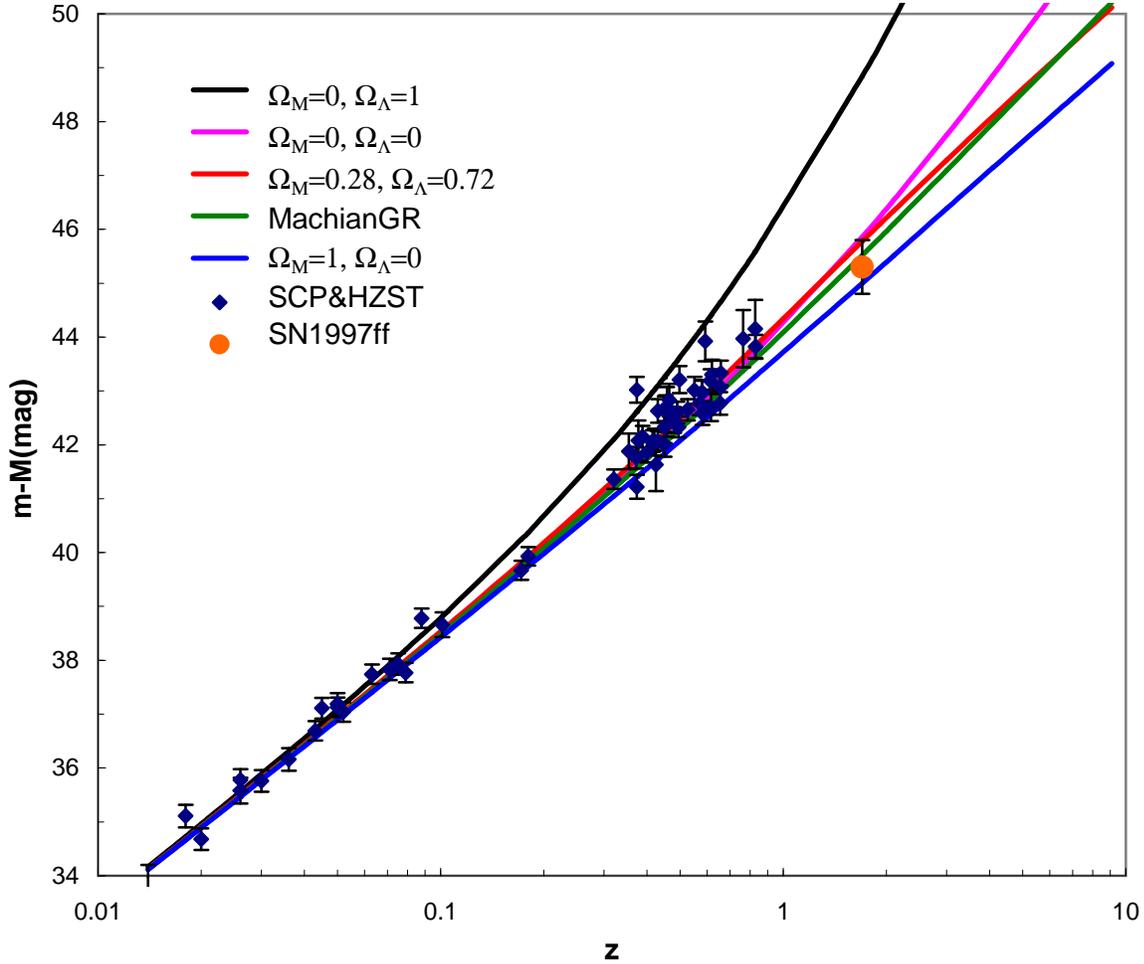}}
  \caption{Hubble diagram showing SNIa results
  \label{fig:supernova}}
\end{figure}

The observational dataset is illustrated in
Figure~\ref{fig:supernova}, together with distance/redshift plots
for a number of alternative cosmologies with varying
$\Omega_\Lambda$ and $\Omega_M$. Results from the various studies
appear to provide unambiguous evidence that the Universe is
accelerating, at least in comparison to a cosmological model
dominated by $ \Omega_M$. The best fit, based on a model with
$\Omega_{Tot} = 1$, is a Universe with $\Omega_\Lambda=0.72$ and
$\Omega_M=0.28$.  These values also give an age for the Universe
of 15 Gyr, which is in accordance with estimates from other dating
methods.

\subsubsection{Cosmic Microwave Background} Recent observations of
the CMB \cite{efstathiou:2000, white:2000, deBernardis:2000,
deBernardis:2001} have measured a peak in the power spectrum at
$\ell\sim 210$.  This provides strong evidence for a flat
Universe, with $\Omega_\Lambda + \Omega_M=1$. The position and
amplitude of subsequent peaks at $\ell\sim 540, 840$ are
consistent with a cosmological model having $\Omega_\Lambda=0.7$
and $\Omega_M=0.3$, and a baryon energy density of
$\Omega_bh^2=0.022$, although there is considerable degeneracy
between  $\Omega_\Lambda$ and $\Omega_M$.

\subsubsection{SN 1997ff} 
\begin{figure}
  \centering
  \scalebox{1}{\includegraphics{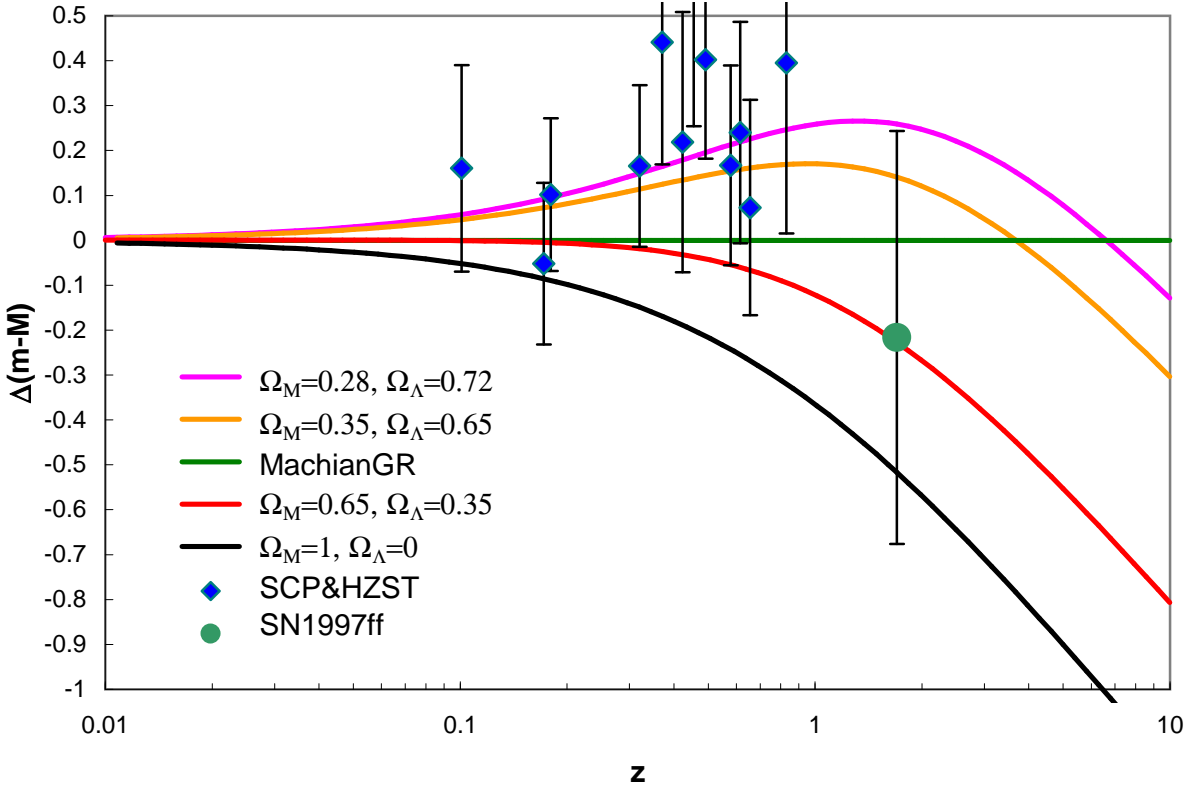}}
  \caption{Hubble diagram showing SNe Ia results plotted against various
  cosmological models
  \label{fig:hubble}}
\end{figure}

Measurements of the luminosity-redshift
relationship for $z<1$ merely indicate a deviation from a standard
matter-dominated unverse, and are not able to distinguish between
various alternative cosmological models. In order to narrow down
the range of possibilities it is necessary to extend observations
to include supernovae with redshifts significantly greater than
one. The serendipitous discovery of SN 1997ff
\cite{reiss:SN1997ff}, the farthest supernova currently known, has
provided the first opportunity to examine the redshift-distance
relationship for cosmological bodies with ages of the order of 10
Gyr. Photometric measurements of this supernova show that it has a
redshift: $z=1.7\pm0.1$, and a distance modulus with a likelihood
function centered around $m-M=45.3$, and 95\% confidence limits of
$+0.5$ and $-1.5$. Figure~\ref{fig:hubble} shows the redshift and
distance data from the SNe Ia measurement programmes, plotted
together with the predictions from a number of possible
cosmological models.  The redshift and distance for SN1997ff is
shown in relation to these models.  The data and models in this
figure are plotted as a delta from a linearly expanding Universe.

Although it is not valid to base any detailed analysis on a single
data point, certain qualitative conclusions can be drawn from the
SN1997ff result.  The magnitude likelihood function still allows
for the possibility of either a conventional decelerating Universe
with $\Omega_M=1$, or a Universe containing dark energy with the
parameters that provided the best fit based on previous SN Ia
data, i.e. $\Omega_\Lambda=0.28, \Omega_M=0.72$.  However, the
$\Omega_M=1$ case does not fit with data from supernovae having
$0.5<z<0.8$, and there is only a low probability of a dark energy
scenario being consistent with the SN1997ff result if
$\Omega_\Lambda=0.28$. A better fit with the SN1997ff data can be
obtained with $\Omega_\Lambda=0.35$ and $\Omega_M=0.65$.  However,
these parameter values give an age for the Universe of about 12.5
Gyr, which is inconsistent with other independent estimates.

\par Figure~\ref{fig:hubble} also shows a plot for a flat, empty
Universe, with $\Omega_\Lambda=0, \Omega_M=0, \Omega_k=0$.
Substituting these values in (\ref{lumdist}) gives

\begin{eqnarray*}
 d_L  & = & c H_0^{-1}(1+z) \int_0^z(1+z)^{-1}dz'\rbrace \\
   &=& c H_0^{-1}(1+z) \cdot log_N(1+z)
\end{eqnarray*}

From Equation \ref{lookback}, the age of this Universe is given by

\begin{eqnarray*}
 H_0 t_0  & = & \int_0^\infty  {(1 + z)^{ -2}} dz\\
  \Rightarrow t_0&=& 1/H_0
\end{eqnarray*}

This describes a linearly expanding Universe, with an age of 15.6
Gyr, based on a current value for $H_0$ of $63km s^{-1} Mpc^{-1}$.
This model gives a predicted distance modulus of $m-M=45.5$ at a
redshift $z=1.7$, which coincides with the highest probability
region of the magnitude likelihood function for SN1997ff.  This
model also exhibits the observed dimming behavior in comparison
with a matter-dominated decelerating Universe for $0.5<z<0.8$,
albeit to a lesser extent than models with $\Lambda\neq0$. Clearly
we do not live in an empty Universe, so at first sight this model
would appear to be of little use in describing a real Universe.
However, we shall see in the following section that it is possible
to construct an alternative cosmology that mimics the behaviour of
an empty Universe, and yet has an energy density equivalent to the
critical value without the need for a non-zero cosmological
constant.

It is of interest to note at this point another apparent fine
tuning problem in addition to the problems discussed in Section
\ref{sec:bigbang}. The currently favored values for
$\Omega_\Lambda$ and $\Omega_M$, based on a best fit with
supernova data for $z<0.8$, are also the values which result in an
age for the Universe which is in accordance with other independent
estimates. These values are also very close to the values required
for $H_0=1/t$ (see Figure \ref{fig:age}). This may be purely
coincidence, but if not, some additional fine tuning mechanism
would be required to cause the ratio $\Omega_\Lambda/\Omega_M$ to
take on the precise value necessary to mimic the behaviour of a
linearly expanding Universe in which $H_0=1/t$.

\section{Machian General Relativity}
\subsection{Rationale for change}
Before examining possible modifications to General Relativity that
might give rise to alternative cosmologies, it is perhaps
worthwhile considering the rational for changing what is, by any
standards, a highly successful theory that has been extensively
verified by numerous experiments.  Arguably, from a theoretical
perspective,  the most compelling reason for seeking an
alternative to the existing Einstein Field Equations is that
General Relativity stands alone amongst gauge field theories in
not being scale-invariant.  This fact has motivated attempts by
many workers to develop a scale-invariant theory of gravity. These
have taken various forms, most notably the scalar-tensor theories
first proposed by Brans and Dicke \cite{dicke:mach}. Other
formulations have also been proposed, including the
scale-covariant theory of Canuto \cite{canuto:1977,canuto:1978},
and the conformal gravity  of Mannheim \cite{mannheim:1990}. The
scale dependency inherent in General Relativity leads to an
apparently fundamental scale, defined by the Planck units:

\paragraph{Planck Length}
\begin{equation}\label{Lplanck}
{\rm  }L_P  = \sqrt {\frac{{\hbar G}}{{c^3 }}} \simeq 10^{-35}m
\end{equation}

\paragraph{Planck Time}
\begin{equation}\label{Tplanck}
 T_P  = \frac{{L_P }}{c} \simeq 10^{-43}s
\end{equation}

\paragraph{Planck Mass}

\begin{equation}\label{Mplanck}
M_P  = \sqrt {\frac{{\hbar c}}{G}} \simeq 10^{-8}kg
\end{equation}

No theory of Quantum Gravity is yet able to explain the wide
disparity between the Planck scale and the atomic scales that
govern the everyday world we inhabit.  Whilst the magnitude of the
Planck length and time scales, at $\simeq 10^{-20}$ of the
corresponding atomic length and time scales, are just about
reconcilable with the concept of an evolving quantum Universe, it
is difficult to account for the fact that the Planck mass is
$\simeq 10^{20}$ times larger than the proton mass.

\par Cosmology presents another set of reasons for considering
modifications to General Relativity.  In Section \ref{sec:bigbang}
the problems associated with Big Bang cosmology were reviewed.
Whilst these can be explained to an extent by concepts such as
inflation and dark energy, these solutions actually raise as many
problems as they solve.  The Big Bang model is a direct
consequence of Friedman-Robinson-Walker cosmology, which in turn
is derived from the field equations of General Relativity.  If a
more elegant solution is to be found to the problems currently
associated with the Big Bang model, then it may well be necessary
to address this at source.

\par Finally, it can be argued that the presently accepted
formulation of General Relativity leads to problems at small
scales and high energies, which can only be resolved by an as-yet
undiscovered theory of Quantum Gravity.  The most significant
manifestation of this problem is the apparent inevitability of
singularities associated with Black Holes and the Planck epoch at
the origin of the Universe.

\subsection{Mach's Principle} Mach's Principle, in its most basic
form, asserts that the inertia experienced by a body results from
the combined gravitational effects of all the matter in the
Universe acting on it.  Although a great admirer of Mach, Einstein
was never entirely certain whether General Relativity incorporated
Mach's Principle.  Indeed, the issue is still the subject of
continued debate even today (see \cite{barbour:mach} for example).
A stronger version of Mach's Principle can be formulated, which
states that the inertial mass energy of a matter particle is equal
and opposite to the sum of the gravitational potential energy
between the particle and all other matter in the Universe, such
that:
\begin{eqnarray}
    mc^2  & = & - \sum\limits_N {\frac{{Gm.m}}{r}}  \\
    &= &-4\pi \alpha Gm\int\limits_0^R {\rho (r)r^2
    .\frac{1}{r}dr}\label{eqn:mach}
\end{eqnarray}

where $R \equiv c/H$ is the gravitational radius of the Universe,
and $\alpha$ is a dimensionless constant.  In the case of a
homogeneous and isotropic matter distribution this becomes

\begin{equation}
    mc^2  = -2\pi \alpha Gm \bar \rho R^2 \label{eqn:mach2}
\end{equation}

where $ \bar \rho $ is the average matter density of the Universe.

\par Observational evidence suggests that the relationship  \( {{G\rho
_0 } \mathord{\left/
 {\vphantom {{G\rho _0 } {H_0^2  \simeq 1}}} \right.
 \kern-\nulldelimiterspace} {H_0^2  \simeq 1}}
\) is valid to a reasonable degree of precision in the present
epoch.  It would be particularly satisfying if this relationship
were to be found to be true, as it would tie in with the concept
that the Universe is `a free lunch', i.e. all the matter in the
Universe could be created out of nothing, with a zero net energy.
However, it can readily be seen from (\ref{eqn:mach2}) that as the
Universe continues to expand, the energy arising from
gravitational attraction will ultimately tend towards zero.
Conversely, gravitational energy will become infinite at $t=0$,
the initial singularity. Since there is no suggestion that the
rest mass energy associated with the matter in the Universe
changes over time (unless one is considering the Steady State
Theory), it would appear that this neat zero energy condition in
the present era is just a coincidence.

\par It can also be seen that the integral in (\ref{eqn:mach}) will tend towards
infinity in regions of space where the gravitational field
strength becomes very high, e.g. in the vicinity of a Black Hole.
In order for Mach's principle to apply not just in the present
epoch, but for all time and over all space, we would have to
forego the concept of a fixed Newtonian gravitational constant.
Accordingly, we use (\ref{eqn:mach2}) to derive an expression for
$G$
\begin{equation}\label{eqn:G}
G = \frac{\alpha c^2}{{2\pi R^2 \bar \rho }}
\end{equation}

\subsection{Einstein field equations revisited}
Arguably Einstein's key insight in his original formulation of the
field equations of General Relativity was the postulate that
space-time could be curved by the presence of energy, together
with the principle of general covariance. This leads to the
construction of the Einstein curvature tensor from components of
the Riemann tensor:
\[
G_{\alpha \beta }  = R_{\alpha \beta }  + g_{\alpha \beta } {R
\mathord{\left/
 {\vphantom {R 2}} \right.
 \kern-\nulldelimiterspace} 2}
\]
The relationship between the Einstein curvature tensor and the
stress-energy tensor is then given by:
\[
G_{\mu \nu }  = KT_{\mu \nu }
\]
where ${K}$ is a scalar `constant' that determines the extent to
which a given amount of stress-energy is able to curve space-time.
The requirement that this should produce results that are
compatible with Newtonian gravity, in the limit of weak, slowly
varying gravitational fields, leads to the formulation of the full
Einstein field equation:
\begin{equation}\label{einstein}
G_{\mu \nu }  = \frac{{8\pi G}} {{c^4 }}T_{\mu \nu }
\end{equation}

The traditional  cosmological constant term $\Lambda$ has been
omitted from this equation.  It will be shown that the observed
dynamical behaviour of the Universe can in fact be explained with
$\Lambda=0$.
\par Whilst the inclusion of $G$ in
the Einstein equation has the desired effect of achieving
compatibility with Newtonian gravity, it could be said that in
`writing in' the Newtonian gravitational constant in this way,
Einstein missed the opportunity to make a clean break with
Newtonian gravity and to offer a more fundamental derivation for
the observed scale of space-time curvature generated by
stress-energy.
\par We are therefore looking for a formulation for the gravitational field
equations that preserves the essential form of the curvature and
stress-energy tensors, and yet incorporates Mach's principle at a
fundamental level.  This suggests the following postulate

\begin{postulate}
The curvature of spacetime in a given region of the Universe,
relative to a surrounding region, is proportional to the ratio of
the stress-energy density in that region to the stress-energy
density in the surrounding region.
\end{postulate}

In this context, the Universe is defined as being the volume of
spacetime encompassed by a 3-sphere with a radius equal to the
speed of light times the age of the Universe. We now take the key
step of asserting that the Machian energy condition should be
valid for any spacetime coordinate. Replacing the gravitational
constant in (\ref{einstein}) with the expression for $G$ in
(\ref{eqn:G}), leads to a redefinition of the gravitational field
equation

\begin{equation}\label{omega1}
G_{\mu \nu }  = \frac{3}{{R^2 \bar \rho c^2 }}T_{\mu \nu }
\end{equation}

where $\bar \rho$ is the effective gravitational mass density,
defined as

\begin{equation}\label{rhobar}
\bar \rho R^2  = 2\int\limits_0^R {\rho (r)r^2\frac{1}{r}dr}
\end{equation}

The geometrical constant $\alpha$ in (\ref{eqn:G}) must take the
value $3/4$ for compatibility with Newtonian gravity and GR.

\subsection{Cosmic dynamics}\label{sec:dynamics}
The components of the Einstein tensor are derived from the
Robertson-Walker metric in the usual way to give

\begin{eqnarray}\label{g00}
    G_{00} & = & \frac{{3\dot a^2 }}{{a^2 c^2 }} + \frac{{3k}}{{a^2
    }}\\
    G_{11} & = & - \frac{{k + {{2a\ddot a} \mathord{\left/
 {\vphantom {{2a\ddot a} c}} \right.
 \kern-\nulldelimiterspace} c}^2  + {{\dot a^2 } \mathord{\left/
 {\vphantom {{\dot a^2 } {c^2 }}} \right.
 \kern-\nulldelimiterspace} {c^2 }}}}{{1 - k\sigma ^2 }} \label{g11}
\end{eqnarray}

The equations of motion are derived by combining
(\ref{g00},\ref{g11}) with the corresponding 00 and 11 components
of the modified field equation (\ref{omega1}) to give

\begin{eqnarray}\label{d2}
  \frac{{\dot a^2 }}{{a^2 }} + \frac{{kc^2 }}{{a^2 }} & = & \frac{{c^2 \rho }}{{a^2\bar
  \rho
  }}\\
 - \frac{{2\ddot a}}{a} - \frac{{\dot a^2 }}{{a^2 }} -
\frac{{kc^2 }}{{a^2 }}
 & = & \frac{{p}}{{\bar \rho a^2 }} \label{d3}
\end{eqnarray}

Since $p$ is small in the present epoch, (\ref{d3}) becomes

\begin{equation}
\frac{{2\ddot a}}{a} + \frac{{\dot a^2 }}{{a^2 }} + \frac{{kc^2
}}{{a^2 }} = 0
\end{equation}

and (\ref{d2}) simplifies to

\begin{equation}\label{expand}
\dot a = c\sqrt {\frac{\rho }{{\bar \rho }} - k}
\end{equation}

From this equation it can be seen that when $\rho={\bar \rho }$,
and the curvature  $k=1$, the result is a pseudo-static solution
similar to the Einstein-DeSitter model, in that it has zero net
energy and $\dot a=0$. The solution is pseudo-static in that the
Universe is only flat and static in a cosmological reference
frame. In regions of matter concentration where $\rho>{\bar \rho
}$, $k$ will effectively appear to be zero and the Universe will
therefore appear to be expanding to an observer in this reference
frame, with its horizon receding at the speed of light.
\par This equation also embodies the negative feedback mechanism
that ensures that $\rho_{mat}$ will always be equal to
$\rho_{grav}$.  If at any point $\rho_{mat}$ should exceed
$\rho_{grav}$ then this will lead to a positive $\dot a$, which
will tend to drive $\Omega\rightarrow1$.  The converse will apply
if $\rho_{mat}$ should fall below $\rho_{grav}$.

\subsection{Energy-momentum conservation}
One of the defining features of the Einstein gravitational field
equation (\ref{einstein}) is the fact that both sides of the
equation are symmetric, divergence-free, second rank tensors. The
stress-energy tensor $T_{\mu\nu}$ embodies the laws of energy and
momentum conservation, such that ${T^{\beta\alpha}}_{;\alpha}=0$.
However, at fist sight the RHS of the modified field equation in
(\ref{omega1}) would appear not to be divergence-free in that the
expression that replaces the gravitational constant $G$ seems to
be time dependent, i.e.
\begin{equation}\label{eqn:divergence}
  \frac{\partial }{{\partial x^0 }}\left(
{\frac{3}{{R^2 \bar \rho c^2 }}} \right) \ne 0
\end{equation}

It would seem that either we have to forego the divergence-free
nature of the original Einstein equation, or we have to `adjust'
the new equation in some way in order to retain this desirable
property. The latter approach was used in the Brans-Dicke
scalar-tensor theory, and subsequently in the Canuto
scale-covariant theory. Although these fixes solved the immediate
problem by restoring the zero-divergence property, this was done
at the expense of the overall elegance of the solution, and
ultimately its ability to make any useful predictions.
\par How, then, are we to resolve this issue whilst still
retaining the logical Machian form of the revised equation
(\ref{omega1})?  The proposed solution turns out to be both
simple, and yet far reaching in terms of its potential impact. It
is to redefine the nature of time within General Relativity and
Quantum Mechanics.

\subsection{Scale time}\label{sec:time}
In order to preserve the principle of general covariance, and
still retain the features of the gravitational field equations in
(\ref{omega1}) we introduce the concept of scale time, $\tau$,
defined as
\begin{equation}\label{tau}
    \tau= \frac{1}{a}
\end{equation}

where $a$ is the scale factor. Scale time is closely related to
conformal time $\eta$, with
\begin{equation}
\frac{{d\eta }}{{dt}} = \frac{1}{a} = \tau
\end{equation}

If the Machian gravitational field equations are to exhibit the
desired momentum conservation properties, we must assert that
scale time is in fact the correct definition of time to use in
General Relativity. Accordingly, we substitute for $R\equiv ct$ in
(\ref{eqn:divergence}) with $cn\tau$, where $n$ is the time in
atomic time units.  Since $n\propto a = 1/\tau$, we see that the
expression in (\ref{eqn:divergence}) is invariant with respect to
$x^0\equiv \tau$, and hence possesses zero divergence.

\par The concept of scale time is perhaps best visualised by
considering the Universe to have the topology of a Euclidean
3-sphere, in which spatial position is defined in the usual way by
three angles in spherical polar coordinates, and time corresponds
to the radius of the sphere.  In this model, the concepts of past,
present and future do not have any real meaning. Time is merely a
coordinate that determines the volume of spacetime currently
occupied by a particle or field. This is in contrast to the
conventional concept of spacetime embodied in a Lorentz-de Sitter
metric, in which time is perceived to flow from a well defined
origin in the past towards an infinite future. Scale time, as
defined here, is essentially similar to the concept of imaginary
time used in some descriptions of Quantum Gravity.

\par This leads to a somewhat
bizarre picture of time and space. From the perspective of an
observer in the cosmological reference frame the Universe would
appear to be static and closed. Any concentrations of matter in
the Universe would appear to be shrinking in size. For an
observer, such as ourselves, linked to the atomic reference frame,
the Universe will appear to be flat and expanding, with the
horizon receding at the speed of light. Our perception of time as
a continuum is an illusion, and the time co-ordinate that we are
used to is perhaps better described as subjective time.  For an
observer in some intermediate reference frame, for example a
photon that was emitted at some time in the past, the picture will
be a mixture of the two scenarios described above: the Universe as
a whole will appear to be expanding, but matter particles will
appear to be contracting. (This notion is not entirely novel;
something similar was proposed by Jeans in 1931
\cite{jeans:1931}).
\par The concept of two kinds of time - cosmological time and
atomic time - is similar in some respects to the dual timescales
postulated by Milne in his kinematic theory of gravity
\cite{milne:1938}.  The important difference is that in the
Machian model, only scale time is consistent with General
Relativity.  All other timeframes are in a sense measurements of
subjective or emergent properties linked to a particular physical
reference frame - in our case, one based on atomic matter.  Such
reference frames are not in any way unique, and one could readily
conceive of a alternative reference frame based, for example, on
photon time.

\section{Consequences}\label{sec:consequences}
In this section we examine the implications of the cosmological
model described in Section 2 in terms of its ability to solve the
Big Bang problems posed in Section \ref{sec:bigbang}, and to offer
a reasonable explanation for the observed behaviour of the
Universe. Throughout most of this section we shall be evaluating
the effects of the model from the perspective of an observer in
the atomic reference frame. Under such circumstances it will be
appropriate to use the time-varying gravitational `constant'
formulation, with $G(t)\propto t$, where $t$ is the conventional
(i.e. subjective) time.

\subsection{Solving the Big Bang problems}

\subsubsection{The flatness problem}
The dynamical equations in Section 2.5 clearly show that, for the
Universe as a whole, the mean energy density is maintained at the
critical value by a form of negative feedback mechanism.  Under
such circumstances, any small deviation of $\Omega$ from unity
would result in an apparent acceleration or deceleration of the
expansion rate so as to bring the system back to its equilibrium
state.  The critical energy density of the Universe is given by

\begin{equation}
  \rho _c = \frac{{3H^2 }}{{8\pi G}}
\end{equation}

Using the expression for $G$ in (\ref{eqn:G}) we find that
\begin{equation}
{{\bar \rho } \mathord{\left/
 {\vphantom {{\bar \rho } {\rho _c }}} \right.
 \kern-\nulldelimiterspace} {\rho _c }} = \Omega  = 1
\end{equation}

Recognizing the fact that Postulate 1 inevitably results in a
Universe where $\Omega=1$, we shall use the terms `Omega model' or
`Omega paradigm' as a shorthand for referring to this formulation
in the following discussion.

\subsubsection{The horizon problem}
The Omega model requires the Universe to be spatially closed, with
the topology of a 3-sphere. Because the expansion rate is
constant, such that $H=1/t$, the horizon distance will always be
equal to the radius of the 3-sphere that defines the observable
Universe.  As a result of this coincidence of horizon distance and
the radius of the observable Universe, all regions within the
Universe will have been in causal contact with each other at some
point in time.  This accounts for the observed homogeneity of the
Universe, and provides an elegant solution to the horizon problem.

The smoothness problem, i.e. accounting for the perturbations in
matter density that give rise to structure formation, is not
explained directly by the Omega model. As will become apparent in
Section 4, the model suggests alternative scenarios for the
earliest stages of cosmic evolution that may involve a boson
$\rightarrow$ fermion phase transition. Such transitions would
result in a Universe of virtually absolute uniformity.  The
variations in matter density that are required as a prerequisite
for galaxy structure formation must therefore arise from
statistical decay processes.

\subsubsection{The cosmological constant problem}
This problem is resolved in the Omega model by removing the
requirement for a cosmological constant altogether.  It was shown
in Section \ref{sec:observations} that the experimental evidence
for the apparent acceleration in the expansion of the Universe is
consistent with a linearly expanding Universe model. In Section
\ref{sec:dynamics} we saw that the modified field equations of
Machian GR give rise to just such a linear expansion. Since there
is no theoretical or experimental requirement for a non-zero
$\Lambda$, it can validly be omitted from the gravitational field
equations. Hence the problem of how to explain a small, but
non-zero, $\Lambda$ disappears.

\subsection{The Large Number Hypothesis} A dimensionless quantity
known as gravitational structure constant can be defined as the
ratio of the electrostatic forces between two adjacent charged
particles, e.g. protons, to the gravitational force between the
particles.

\begin{equation}\label{alphaG}
\alpha _G  =  \frac{{Gm_p ^2 }}{{\hbar c}}{\rm (} \approx {\rm
5}{\rm .9} \times {\rm 10}^{{\rm  - 39}} )
\end{equation}

Standard cosmological theories provide no obvious explanation for
such a vast disparity between the forces of gravity and
electromagnetism.  In 1938 Dirac  \cite{dirac:1938} noted that the
dimensionless quantity \( {1 \mathord{\left/
 {\vphantom {1 {\alpha _G }}} \right.
 \kern-\nulldelimiterspace} {\alpha _G }}
\) was approximately equal to the present age of the Universe
measured in atomic time units (where 1 atomic time unit = \(
{\hbar  \mathord{\left/
 {\vphantom {\hbar  {m_p c^2  \approx }}} \right.
 \kern-\nulldelimiterspace} {m_p c^2  \approx }}10^{ - 24}
\) secs).  If this relationship were to be valid for all epochs
then this implies that \( {1 \mathord{\left/
 {\vphantom {1 {\alpha _G }}} \right.
 \kern-\nulldelimiterspace} {\alpha _G }}
\) must be proportional to the age of the Universe, and therefore
that \( G\left( t \right) \propto {1 \mathord{\left/
 {\vphantom {1 t}} \right.
 \kern-\nulldelimiterspace} t}
\).  This postulate formed the basis of Dirac's Large Number
Hypothesis (LNH), which has subsequently provided the inspiration
for a number of alternative cosmological theories. (It is worth
noting that this formulation of the LNH is equivalent to the
expression \( {{G\rho _0 } \mathord{\left/
 {\vphantom {{G\rho _0 } {H_0^2  \simeq 1}}} \right.
 \kern-\nulldelimiterspace} {H_0^2  \simeq 1}}
\) of Mach's principle).

\par Clearly, since Mach's principle has been used to construct
Postulate 1, which in turn forms the basis for formulating the
revised gravitational field equations, the Omega model will
inherently embody the strong version of Mach's principle.
Specifically, the rest mass energy of all matter in the Universe
will be equal and opposite to its mutual gravitational potential
energy, such that the sum of these energies is equal to zero.

In looking at some of the implications of this relationship for
observers in the atomic reference frame it is helpful to express
the mean gravitational energy density of the Universe in terms of
the baryon number $N$, and the mean baryon mass, which we shall
take to be the proton mass $m_p$.  (Note that this implies, but
does not require, that any missing mass in the Universe is
baryonic in nature rather than in the form of other more exotic
entities).
\begin{equation}
    \bar \rho  = \frac{{3Nm_p }}{{4\pi R^3 }}
\end{equation}
The expression for $G$ in (\ref{eqn:G}) can therefore be written
as
\begin{equation}\label{Gsub3}
G = \frac{{R(t)c^2 }}{{2Nm_p }}
\end{equation}

If we now substitute for $G$ in equation (\ref{alphaG}) for the
gravitational structure constant $\alpha_G$, we find that

\begin{equation}
    \alpha _G  = \frac{{R(t)cm_p }}{{N\hbar }} \label{eqn:alphag1}
\end{equation}
where $R(t)$ is the apparent radius of the Universe in the atomic
reference frame, at subjective time $t$.  From this it can be seen
that $\alpha_G \propto t$ in our reference frame, i.e. the
strength of the gravitational interaction between particles will
increase over time in relation to their mutual electromagnetic
forces.  Combining (\ref{eqn:alphag1}) with the expression for
atomic time we find that

\begin{equation}
\alpha _G  = \frac{n}{N}
\end{equation}

where $n$ is the time in atomic time units. Although this very
simple result may at first seem somewhat surprising, it is perhaps
to be expected, since the baryon number $N$ is one of the few
dimensionless quantities to occur naturally in cosmology. (The
fact that $1/\alpha_G \approx n$ today is purely a coincidence).
The implications of the time dependence of $\alpha_G $ for
$t>>t_0$ will be examined in Section \ref{sec:predictions}.

\subsection{Planck Units}
We shall now examine the effects of recasting the expressions for
the Planck units using the formula for $G$ given in (\ref{Gsub3}),
and the de Broglie wavelength of a proton given by \( R \approx
\lambda _p  \approx {\hbar  \mathord{\left/
 {\vphantom {\hbar  {m_p c}}} \right.
 \kern-\nulldelimiterspace} {m_p c}}
\).

\paragraph{Planck Length}
Clearly with $G(t) \propto t$ the quantity known as the Planck
Length in (\ref{Lplanck}) will itself be a function of time such
that \( L_P (t) \propto \sqrt t \).  Substituting for $G$ using
(\ref{Gsub3}), and the expression for the atomic time unit, gives

\begin{equation}
  L_P = \lambda_p \sqrt{\frac{n}{N}}
\end{equation}

where $n$ is the time expressed in atomic time units, and $N$ is
the baryon number of the Universe. It is interesting to note that
at a time $n=N$ the Planck Length will have grown to a size such
that $L_P= \lambda_p$.  (Or conversely, in the cosmological frame,
the proton wavelength will have shrunk below the Planck Length).
The implications of this equality for the ultimate fate of the
Universe will be revisited in Section \ref{sec:predictions}.

\paragraph{Planck Time}
A similar set of expressions can be derived for the quantity known
as Planck Time in (\ref{Tplanck}), to give
\begin{equation}
T_P  = \frac{\lambda_p}{c} \sqrt{\frac{n}{N}}
\end{equation}

\paragraph{Planck Mass}
From Equation (\ref{Mplanck}) it is evident that the Planck Mass
\( M_p (t) \propto t^{ - {\textstyle{1 \over 2}}} \). Again,
substituting for $G$ using (\ref{Gsub3}), with \( R = \lambda _p
\) we find
\begin{equation}
M_P  = m_p \sqrt {\frac{N}{n}}
\end{equation}

And when $n=N$ we see that \( M_P  = m_p \).  It is easy to verify
that these expressions lead to the correct present day values for
the Planck units by inserting appropriate values for the proton
mass and radius, and the current time in atomic time units.

\par Based on this analysis, the conclusion we must reach is that Planck Units
do not represent a fundamental measurement scale that becomes
relevant during the birth of the Universe and governs the realm of
Quantum Gravity. Rather, they are scale factor dependent
quantities which may shed some light on the behaviour of the
Universe in its dying moments.

\subsection{Cosmological observations}
We have already seen in Section \ref{sec:dynamics} that a linearly
expanding Universe, such as the one that would evolve from the
modified gravitational field equations, will give rise to the
currently observed Universe, with:
\begin{itemize}
  \item Hubble parameter $=63 km s^{-1}Mpc^{-1}$ for a Universe of
  age $\sim 15 Gyr$.
  \item Critical density $\Omega=1$
  \item Apparent acceleration relative to a Universe with
  $\Omega_M=1$
\end{itemize}

\subsection{Variations in the gravitational constant}
The prediction that $G(t) \propto a$, where $a$ is the scale
factor, is another principal feature of the Omega model. However,
it will not be possible to detect any variation in \( {\dot G/G}
\) by, for example, measuring changes in planetary orbits within
the solar system using radar ranging techniques, since time
measured by any atomic or gravitational clock will be changing at
the same rate as the distance to be measured.  Suppose at time
$t_0$ and scale factor $a_0$ the measured distance is
$2r_0=cn\tau_0$, where $\tau_0$ is the period of an atomic clock
and $n$ is the number of clock ticks between the emission of a
radar signal and the reception of its reflection. At some future
time when the scale factor has increased to $a$, the measurement
is repeated.  If $G(t) \propto a$ then the distance to the planet
will have decreased so that $r=r_0a_0/a$.  However, the period of
the atomic clock will also have decreased by the same proportion,
with $\tau=\tau_0a_0/a$.  Consequently, the measured elapsed time
for the radar signal round trip will still be $n$ ticks, i.e.
there will be no apparent change in distance and therefore no
change in gravitational constant.

\par In order to verify that $G$ does vary over cosmological timescales
it will be necessary either to measure its value directly using a
Cavendish type experiment, or to turn to evidence from geophysical
and astrophysical measurements, and from models of galactic
evolution. Since Cavendish experiments can currently only achieve
accuracies of one part in $10^{-6}$, these are not capable of
detecting changes in $G$, which will be of the order of the Hubble
factor, i.e. one part in $10^{-11}$ per year. We must therefore
look to the other sources for indirect evidence of a time-varying
$G$.

\subsection{Varying Speed of Light Cosmology}
A novel cosmology, based on the concept of a varying speed of
light, has been proposed by Albrecht, Barrow, and Magueijo
(\cite{jm:2, jm:3, jm:4, jm:5}).  This is capable of explaining
the Big Bang problems by postulating that, at an epoch
corresponding approximately to the inflationary era in the
standard Big Bang plus inflation model, the speed of light was
much greater than it is today. In the Omega model, the fact that
$da/d\tau=dt/d\tau$ means that the velocity of light, $c$, must be
constant as measured by an observer at scale factor $a$. However,
it is easy to see that if an observer had some means of measuring
conventional time, rather than subjective time derived from scale
time, then they would perceive a steady decrease in the speed of
light with time. In other words, a linearly expanding Universe
consistent with Machian GR is equivalent to a static Universe in
which the speed of light varies.

\section{Predictions}\label{sec:predictions} As with any other
theory, the usefulness of the Omega paradigm hinges on its ability
not only to explain currently observed phenomena, but also to
successfully predict new phenomena which can subsequently be
verified by experiment.  It has already been shown in Section
\ref{sec:consequences} that the Omega model can provide
explanations for some of the problems associated with the standard
Big Bang model that are considerably more economical than other
prevailing theories. Similarly, it has provided a logical
explanation for the presently observed values of a number of
fundamental physical quantities. In this section we use the
underlying features of this model to make several predictions
pertaining to the history and the fate of the Universe.

\subsection{Photon energy conservation}
The concept of scale time, which is an essential component of the
Omega model, necessitates a reappraisal of the way in which the
equivalence principle is applied in General Relativity.
Essentially, the notion of time as a linear flow of events with a
past, present and future, must be replaced by a picture of time as
a fourth dimension of finite extent.  Scale time is a measure of
the proportion of this time dimension that is occupied by a
particle or field.  With this formulation of time it is evident
that there is no preferred `position' in the time dimension, any
more than there is in the three spatial dimensions.  One can, of
course, still define specific reference frames in which to carry
out physical measurements, the most obvious being the atomic
reference frame that we use for most everyday purposes.
\par
Extending the equivalence principle to embrace this concept, its
is clear that the Hamiltonian of a system should be independent of
the choice of time co-ordinate.  This presents little conceptual
difficulty when applied to a system of particles or bound fields.
However, when the principle is extended to photons it is hard to
escape the conclusion that the energy of a photon must be
independent of the reference frame of the observer. In other
words, photons will retain the same energy that they originally
possessed at the time of their emission, whether they are
red-shifted (or blue-shifted) as a result of
\begin{itemize}
  \item Doppler shift with respect to a given observer
  \item climbing out of a gravitational potential
  \item being `stretched' by the Hubble expansion of the Universe
\end{itemize}
The observable quantity that will change in the reference frame of
the observer will be the \emph{power} of the photon.
\par Clearly, if this prediction is valid then it would have fundamental
implications across many fields of physics.  However the effects
should be experimentally verifiable.  In the field of cosmology,
the most obvious place to look for evidence of photon energy
conservation is in the cosmic microwave background (CMB). The
standard theory is based on the assumption that as the Universe
expands, the energy density due to matter decreases as
$R^{-3}(t)$, whereas the energy density due to radiation decreases
as $R^{-4}(t)$ because of the additional energy loss due to the
red-shift.  In the Omega model this remains true when looking at
the spatial energy density. However, if one is carrying out
measurements of photon energy by integrating power measurements
over a period of time that is long in relation to the time span of
the photon wavepacket (i.e. $t>>\lambda/c$), then the energy will
be found to decrease in proportion to $R^{-3}(t)$, as for the
matter case. So, for example, a CMB photon emitted when the
Universe had a temperature of $10^{9 \circ} K$, with a present day
temperature of $2.9^\circ K$, would still retain its initial
energy given by $h\nu  = kT$, rather than  $\sim10^{-9}$ of this
value.

\par It should in theory be possible to measure this effect
experimentally, by analysis of the noise spectrum of the CMB. Most
recent experiments to measure the CMB have focused on obtaining
improved spatial resolution in order to map out temperature
fluctuations, and hence matter distribution, in the early
Universe. This has necessitated integrating microwave power
measurement over timescales that are long enough to achieve an
adequate CMB signal against the background noise level.  An
experiment to measure the CMB noise spectrum will need to have a
much narrower time resolution ($\sim 10^{-6}s$), but conversely,
the measurement can be integrated over a much larger solid angle.
The standard theory predicts that the results of a Fourier
analysis of these measurements will be a constant noise power
level over the entire frequency range, up to the cut-off
determined by the sampling time window. The Omega model predicts
that the noise power spectrum will be in the form of a Gaussian
distribution, with the peak of the distribution corresponding to
the mean CMB photon arrival rate.

\subsection{Primordial nucleosynthesis}\label{sec:nucleosynthesis}
Arguably, one of the few successes of the standard Big Bang model
is its ability to predict the abundances of the light elements
resulting from primordial nucleosynthesis. If the Omega model is
to be of any use then it must also give predictions that are
consistent with observational data. The Omega model implies that
the observed power of the CMB is due to a relatively small number
of energetic photons rather than a very large number of low energy
photons. Assuming that the CMB photons originally had energies of
$\sim 1 MeV$, corresponding to a temperature of $\sim 10^{11
\circ} K$. These have been redshifted to the currently measured
temperature of $2.7^ \circ K$, implying an energy loss of the
order of $10^{11}$ according to the standard theory. Since the
currently observed CMB photon number density, calculated according
to the standard theory, happens to correspond to $\eta_B\equiv
n_B/n_\gamma \sim 10^{11}$ photons per baryon, it follows that
$\eta_B \simeq 1$ in the Omega model.
\par The current CMB photon energy density is approximately
$10^{-3}$ of the observed energy density due to baryonic matter.
Taking these two observations together, this suggests that we are
looking for a nucleosynthesis model that results in one photon per
baryon, with an energy approximately $10^{-3}$ of the proton rest
mass energy.  The most obvious scenario is that of neutron decay,
first proposed by Gamow \cite{gamow:1}.  Initial studies of the
evolution of a cold neutron Universe have been carried out using
numerical simulation models.  These show that the initially cold,
dense, neutron cloud heats up by means of $\beta$ decay to form a
hot proton-neutron-electron plasma at a temperature of $\sim
10^{11  \circ} K$.  At $\sim 10^{9  \circ} K$ a range of fusion
reactions become energetically favorable, and lead to the
formation of deuterium, tritium, helium and other light elements,
as in the standard Big Bang nucleosynthesis models
\cite{wagoner:1967}.  The main differences between the Omega model
and the standard Big Bang is that in the former, photons play a
negligible role in the exchange of energy between particles.  The
fact that the expansion rate is much slower also has a significant
impact in that there is more time for fusion reactions to take
place, and therefore an increased probability of synthesizing
heavier elements than would be the case with the standard model.
Nucleosynthesis in a linearly expanding Universe has also been
studied by Lohiya in \cite{lohiya:1998}.
\par One of the most important consequences of nucleosynthesis in the Omega model is
that the primordial baryonic matter is not able to cool down as
the Universe expands, since there are insufficient photons to
remove the entropy generated by the neutron decay and nuclear
fusion processes.  The primordial hydrogen and helium molecules
will therefore remain in an ionized state indefinitely.  This may
explain why intergalactic gas clouds are currently observed to be
ionized.  Another feature of this model is that the $\beta$ decay
process that causes the primordial universe to heat up will give
rise to scale-invariant differences in temperature, due to the
statistical nature of the neutron decay reaction. Starting from a
perfectly isotropic and homogeneous state, this mechanism is
therefore able to account for the observed scale-invariant
temperature fluctuations in the CMB, which are explained by
inflation in the standard Big Bang model.

\subsection{Black Holes and singularities}
The modified gravitational field equation in (\ref{omega1}) is
only valid where the matter density is homogeneous and isotropic.
In the more general case where $\rho=\rho (r)$, it is necessary to
use the full version of the modified field equation
\begin{equation}
G_{\mu \nu }  = \frac{3}{{c^2 \int\limits_0^R {\rho (r)rdr}
}}T_{\mu \nu }
\end{equation}

From this it can be seen that in regions of high energy density,
such as the vicinity of a Black Hole,  the density integral will
be much larger than its average value for the Universe as a whole.
The resulting curvature due to stress-energy in $T_{\mu \nu } $
will therefore be proportionately less than it would be in regions
of average matter density. As an example of the consequence of
this phenomenon, consider two stars orbiting in close proximity to
a Black Hole.  Under conventional GR, the mutual gravitational
attraction between these two bodies would be given by the standard
Einstein gravitational equation.  With Machian GR, the additional
curvature induced in the metric as a result of the mass of each of
the stars will tend towards zero as they approach the Swartzchild
radius of the Black Hole.  Expressed in terms of a simple
mechanical analogy, we could say that spacetime becomes stretched
to its elastic limit in the vicinity of a Black Hole, to the
extent that the presence of additional mass-energy is not able to
increase curvature any further.

\par A further consequence of Machian GR is that if Postulate(1) holds true under all circumstances,
then it follows that as matter passes through the event horizon of
a Black Hole, it enters a region of spacetime that is effectively
a separate Universe from the one that exists outside the Black
Hole. Under such circumstances, all dimensions are rescaled, and
there is no central singularity.

\subsection{Implications for Quantum Gravity}
The discussion of the scale factor dependency of the gravitational
structure constant $\alpha_G$ and the Planck units leads to the
conclusion that the conditions applying in the very early Machian
GR Universe would bear little resemblance to the scenario
envisaged in the standard Big Bang model.  Specifically, the
Penrose-Hawking singularity theorem \cite{hawking:1970}, which is
predicated on a conventional concept of time and gravity, would no
longer apply under the Omega model.  In fact the proto-Universe
would be a much more benign environment, without any initial
singularity as such. In the absence of an initial singularity,
there is no need to resort to the concept of a cut-off point
occurring at the Planck length and time.  In any case, we have
seen that in the Omega model these quantities are themselves time
dependent, and do not therefore carry the same fundamental
significance that is attached to them in the standard theory. The
existence the singularity at the Planck era is one of the two
principle motivations for the pursuit of a quantum theory of
gravity (the other being the need for the curvature terms in the
gravitational field equation to be quantized in order to be
equivalent to the stress-energy terms). If this factor is removed,
we need to ask whether there is still a need for a theory of
Quantum Gravity, at least in the form currently being sought.

\par In Section \ref{sec:time}, an alternative spacetime structure
was described, which potentially allows processes to occur
'simultaneously', but at different time coordinates.  This has a
number of implications, including the possibility that several
particle 'generations' can coexist within the Universe, each at a
different stage in their time evolution. This might conceivably
provide an explanation for the missing mass problem, which
incidentally is not addressed by the Omega model.  The spacetime
structure also offers the intriguing possibility that processes
operating in the cosmological reference frame, i.e. across the
whole Universe,  can occur in the same time as local processes
occurring in the atomic reference frame.  If this were to be the
case, then it provides a mechanism for explaining the apparent
paradox of action-at-a-distance and non-locality associated with
Quantum Mechanics.

\subsection{The ultimate fate of the Universe}
In Section \ref{sec:consequences} we saw that at a time $n=N$
(where $n$ is the time in atomic time units, and $N$ is the baryon
number of the Universe), the evolution of the Omega Universe
reaches a state at which the Planck length is equal to the Compton
wavelength of the proton, and the Planck mass is equal to the
proton mass. Recalling that the Schwartzchild radius of a black
hole is given by
\begin{equation}
    R_S  = \frac{{2GM}}{{c^2 }}
\end{equation}

and substituting for $G$ using (\ref{Gsub3}), with $R(t)=N
\lambda_p$ and $M=m_p$, we find that
\begin{equation}
     R_S  = \lambda_p
\end{equation}

In other words, the scale factor of the Universe has evolved to
the point where the radius of the proton exceeds the Schwartzchild
radius corresponding to the proton mass.  (The term proton here is
used loosely to refer to whatever state baryonic matter may exist
in the extreme gravitational conditions prevailing at this epoch.
In practice it is more likely that protons and electrons will have
recombined into atomic hydrogen by this stage, which in turn may
have collapsed into neutrons in a reversal of the process
described in \ref{sec:nucleosynthesis} above). At this point the
Universe effectively comes to an end as all protons simultaneously
collapse into micro Black Holes - possibly to give birth to many
more baby Universes according to Smolin in \cite{smolin:1994}.

\bibliographystyle{plain}
\bibliography{MachianGR}

\end{document}